\documentclass{article}

\usepackage{amsmath}
\usepackage{arxiv}
\usepackage[utf8]{inputenc} 
\usepackage[T1]{fontenc}    
\usepackage{hyperref}       
\usepackage{url}            
\usepackage{booktabs}       
\usepackage{amsfonts}       
\usepackage{nicefrac}       
\usepackage{microtype}      
\usepackage{lipsum}
\usepackage{graphicx}
\graphicspath{ {./images/} }

\title{Autoencoder-based Denoising Defense against Adversarial Attacks on Object Detection}

\author{
School of Cybersecurity in Korea University\\
Hacking and Countermeasure Research Lab\\[0.5em]
(Min Geun Song, Gang Min Kim, Woonmin Kim, Yongsik Kim, Jeonghyun Sim, Sangbeom Park, Huy Kang Kim)\\[1em]
}

\begin{document}
\maketitle
\begin{abstract}

Deep learning-based object detection models play a critical role in real-world applications such as autonomous driving and security surveillance systems, yet they remain vulnerable to adversarial examples. In this work, we propose an autoencoder-based denoising defense to recover object detection performance degraded by adversarial perturbations. We conduct adversarial attacks using Perlin noise on vehicle-related images from the COCO dataset, apply a single-layer convolutional autoencoder to remove the perturbations, and evaluate detection performance using YOLOv5. Our experiments demonstrate that adversarial attacks reduce bbox mAP from 0.2890 to 0.1640, representing a 43.3\% performance degradation. After applying the proposed autoencoder defense, bbox mAP improves to 0.1700 (3.7\% recovery) and bbox mAP@50 increases from 0.2780 to 0.3080 (10.8\% improvement). These results indicate that autoencoder-based denoising can provide partial defense against adversarial attacks without requiring model retraining.

\end{abstract}

\\

\textbf{Acknowledgment}

This work was supported by Institute for Information \& Communications Technology Planning \& Evaluation (IITP) grant funded by the Korea government (MSIT) (No. 2020-0-00374, Development of Security Primitives for Unmanned Vehicles).

\section{Introduction}

Deep learning-based convolutional neural networks (CNNs) have achieved remarkable performance across various computer vision tasks, including image classification, object detection, and semantic segmentation~\cite{redmon2016you}. In particular, real-time object detection models such as You Only Look Once (YOLO) serve as critical components in safety-critical applications including autonomous vehicles, security surveillance systems, and robotic vision. However, the discovery that deep learning models are vulnerable to adversarial examples has raised significant security concerns for real-world deployment~\cite{goodfellow2014explaining}.

Adversarial examples are inputs crafted by adding imperceptible perturbations to original images, causing deep learning models to produce incorrect predictions. This phenomenon, first discovered by Szegedy \textit{et al}.~\cite{goodfellow2014explaining}, has led to the development of various attack methods. Goodfellow \textit{et al}.~\cite{goodfellow2014explaining} proposed the Fast Gradient Sign Method (FGSM) based on the linear nature of neural networks, while Madry \textit{et al}.~\cite{madry2017towards} introduced Projected Gradient Descent (PGD), an iterative extension that generates more powerful adversarial examples. These attacks have been extended beyond image classification to object detection and semantic segmentation~\cite{xie2017adversarial}, with demonstrated transferability across different detection models~\cite{wei2018transferable}.

Various defense mechanisms have been proposed to counter adversarial attacks. Adversarial training incorporates adversarial examples into the training data to enhance model robustness~\cite{madry2017towards}, but it is effective only against specific attacks and incurs high computational costs. Input transformation-based defenses apply image transformations such as bit-depth reduction, JPEG compression, and total variance minimization to remove adversarial perturbations~\cite{guo2017countering}. The nature of these transformations makes them difficult for attackers to circumvent. Xie \textit{et al}.~\cite{xie2017mitigating} demonstrated that random resizing and padding at inference time can mitigate adversarial effects without additional training. Autoencoder-based defenses have also gained attention. MagNet ~\cite{meng2017magnet} proposed using autoencoders to learn the manifold of normal data and reform adversarial examples toward this manifold. Liao \textit{et al}.~\cite{liao2018defense} addressed the error amplification effect through a High-level representation Guided Denoiser (HGD), achieving first place in the NIPS adversarial defense competition.

However, existing research has primarily focused on image classification tasks, and the effectiveness of autoencoder-based defenses for object detection has not been systematically evaluated. In particular, quantitative analysis of how denoising-based defenses can recover detection performance in real-time detection models like YOLO remains limited.

In this study, we propose an autoencoder-based denoising defense to recover object detection performance on adversarially perturbed images. We apply adversarial attacks using Perlin noise~\cite{perlin1985image} to vehicle images from the COCO dataset~\cite{lin2014microsoft}, remove the noise through a single-layer convolutional autoencoder, and evaluate detection performance using YOLOv5. 

\section{Related Work}

\subsection{Adversarial Attacks}

Adversarial examples were first discovered by Szegedy \textit{et al}., demonstrating that deep learning models are vulnerable to imperceptible perturbations~\cite{goodfellow2014explaining}. Goodfellow \textit{et al}.~\cite{goodfellow2014explaining} explained this phenomenon as arising from the linearity in high-dimensional spaces rather than the nonlinearity of neural networks, and proposed the Fast Gradient Sign Method (FGSM). FGSM generates adversarial perturbations in a single step using the sign of the loss gradient:
\begin{equation}
x_{adv} = x + \epsilon \cdot \text{sign}(\nabla_x J(\theta, x, y))
\end{equation}
where $\epsilon$ denotes the perturbation magnitude and $J$ represents the loss function.

Madry \textit{et al}.~\cite{madry2017towards} proposed Projected Gradient Descent (PGD), an iterative extension of FGSM that generates stronger adversarial examples. PGD analyzes adversarial robustness from a robust optimization perspective and provides a theoretical foundation for defense against first-order adversaries.

Adversarial attacks on image classification have been extended to object detection and semantic segmentation. Xie \textit{et al}.~\cite{xie2017adversarial} proposed the Dense Adversary Generation (DAG) algorithm to optimize adversarial perturbations across multiple targets (pixels, proposals). Wei \textit{et al}.~\cite{wei2018transferable} introduced the Unified and Efficient Adversary (UEA), generating transferable adversarial examples capable of simultaneously attacking proposal-based models (Faster R-CNN) and regression-based models (SSD, YOLO).

\subsection{Adversarial Defenses}

Defense mechanisms against adversarial attacks can be broadly categorized into adversarial training, input transformation-based defenses, and denoiser-based defenses.

Adversarial training enhances model robustness by incorporating adversarial examples into the training data~\cite{goodfellow2014explaining, madry2017towards}. Madry \textit{et al}.~\cite{madry2017towards} achieved certifiable defense against first-order adversaries by training on PGD-generated adversarial examples. However, this approach is effective only against specific attacks and incurs substantial computational overhead.

Input transformation-based defenses remove adversarial perturbations through image preprocessing. Guo \textit{et al}.~\cite{guo2017countering} demonstrated that transformations such as bit-depth reduction, JPEG compression, total variance minimization, and image quilting are effective against adversarial attacks. Non-differentiable and stochastic transformations are particularly advantageous as they are difficult for attackers to circumvent. Xie \textit{et al}.~\cite{xie2017mitigating} proposed random resizing and padding at inference time to mitigate adversarial effects without additional training.

Autoencoder-based defenses project adversarial examples onto the manifold of normal data. MagNet, proposed by Meng \textit{et al}.~\cite{meng2017magnet}, consists of detectors and reformers, where the reformer uses autoencoders to restore adversarial examples to the vicinity of the normal manifold. However, standard denoisers suffer from the error amplification effect, where residual adversarial noise is amplified in upper layers. Liao \textit{et al}.~\cite{liao2018defense} addressed this issue by proposing the High-level representation Guided Denoiser (HGD), which uses differences in the target model's high-level representations as the loss function.

\subsection{Object Detection}

Object detection is the task of simultaneously predicting the locations and classes of objects within an image. Traditional R-CNN-based models employ a two-stage approach, first generating region proposals and then classifying them. YOLO, proposed by Redmon \textit{et al}.~\cite{redmon2016you}, reformulates object detection as a single regression problem, directly predicting bounding boxes and class probabilities through a unified neural network. YOLO enables real-time processing at 45 frames per second and reduces background errors by leveraging contextual information from the entire image.

\subsection{Dataset and Evaluation Metrics}

The COCO dataset is a large-scale benchmark containing 328,000 images with 2,500,000 instance annotations across 80 object categories~\cite{lin2014microsoft}. With an average of 7.7 objects per image, COCO reflects complex real-world scenes and serves as a standard evaluation dataset for object detection and segmentation. Evaluation metrics include mAP (averaged over IoU thresholds from 0.5 to 0.95), mAP@50 (at IoU threshold 0.5), and mAP@75 (at IoU threshold 0.75).

\subsection{Perlin Noise}

Perlin noise is a procedural noise generation technique proposed by Ken Perlin~\cite{perlin1985image}, widely used for creating natural textures and patterns. Perlin noise exhibits rotational invariance and limited frequency bandwidth, with the octave parameter enabling synthesis of noise at various scales. In this study, we leverage these characteristics to generate adversarial perturbations with natural-looking patterns.

\section{Proposed Method}

\begin{figure}[h]
\centering
\includegraphics[width=0.7\linewidth]{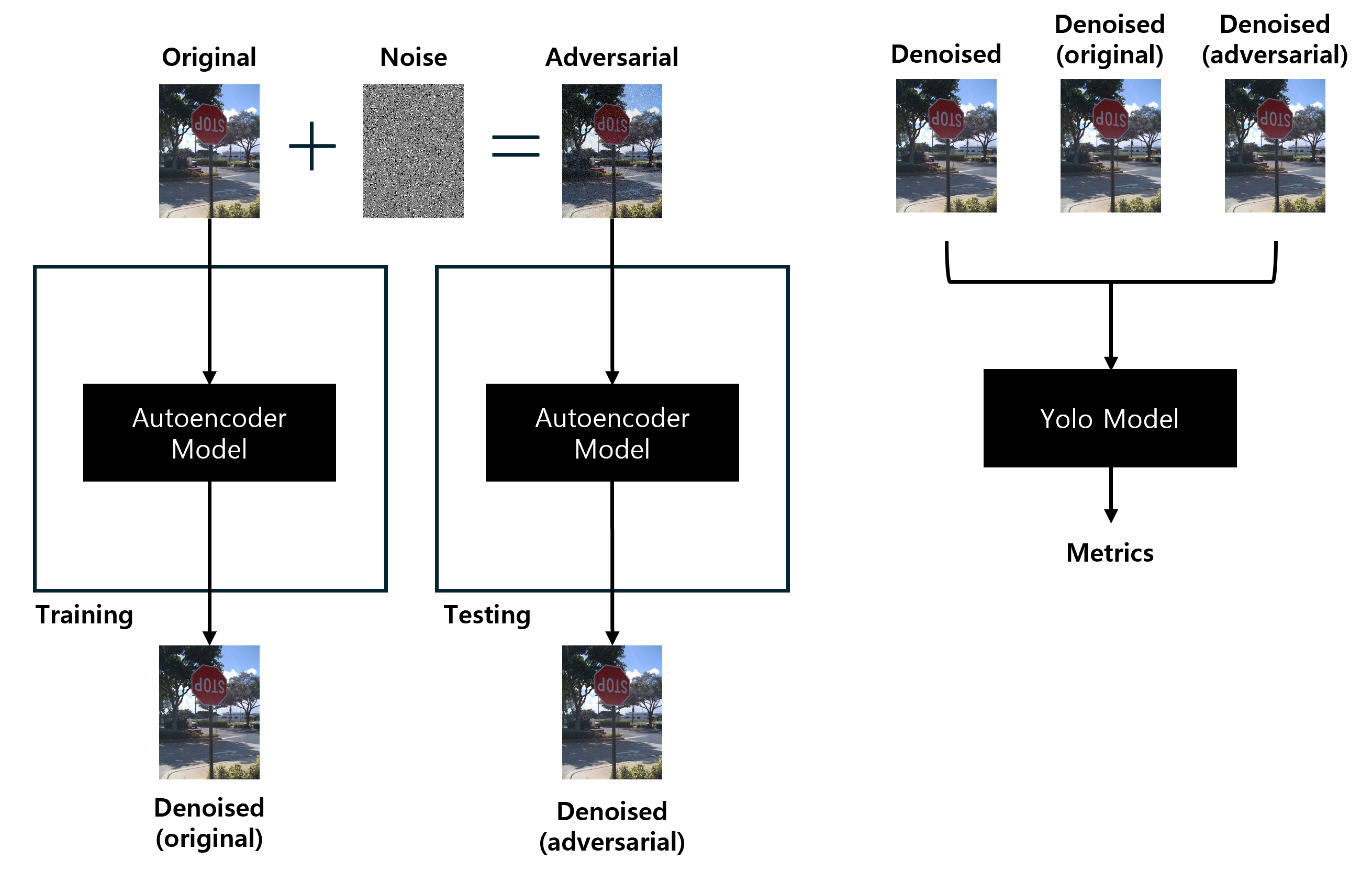}
\caption{Overview of the proposed autoencoder-based denoising defense pipeline. Adversarial noise is added to clean images using Perlin noise, then the autoencoder removes the perturbation before object detection with YOLOv5.}
\label{fig:pipeline}
\end{figure}

Figure \ref{fig:pipeline} illustrates the overall pipeline of our proposed method. We propose an autoencoder-based denoising approach to recover object detection performance on adversarially perturbed images. The proposed method consists of restoring noisy images through an autoencoder before feeding them to the YOLOv5 object detection model.

\subsection{Adversarial Noise Generation}

\begin{figure}[h]
\centering
\includegraphics[width=0.9\linewidth]{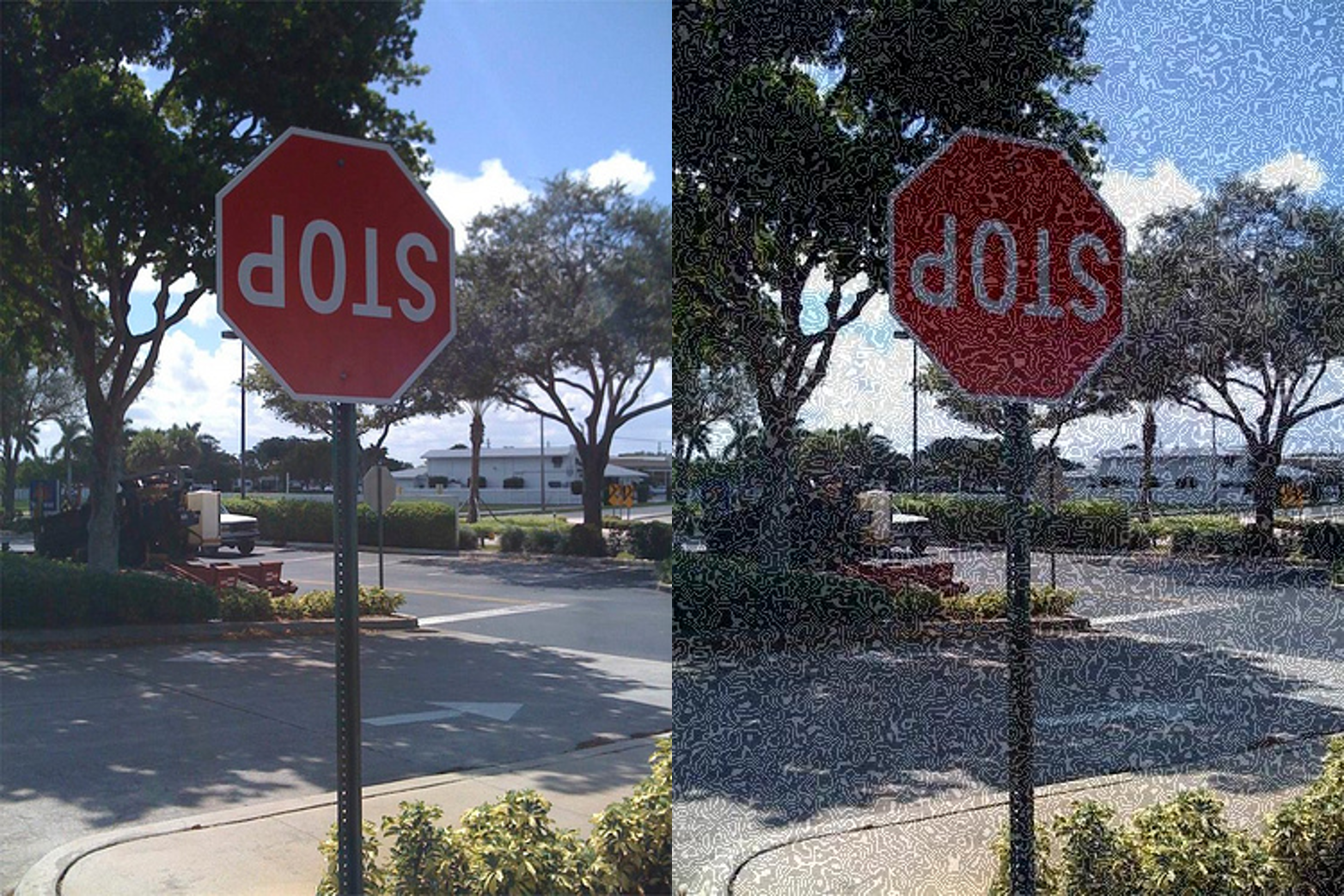}
\caption{Comparison of normal image (left) and adversarial image (right) with Perlin noise perturbation (maximum norm 30, period 30, frequency sine 30, octave 2).}
\label{fig:normal_vs_adversarial}
\end{figure}

Figure \ref{fig:normal_vs_adversarial} shows a comparison between a clean image and its adversarially perturbed counterpart. For adversarial noise generation, we employ Perlin noise-based procedural noise. Perlin noise can generate natural-looking patterns and is configured with the following parameters: maximum norm 30, period 30, frequency sine 30, and octave 2. The maximum norm determines the maximum magnitude of the noise, the period controls the periodicity, the frequency sine adjusts the sine function frequency, and the octave regulates the noise complexity.

The generated noise is applied to original images as follows:

\begin{equation}
I_{adv} = I_{original} + \epsilon \cdot \text{sign}(N_{perlin})
\end{equation}

where $I_{adv}$ denotes the adversarial image, $I_{original}$ the original image, $\epsilon$ the perturbation magnitude, and $N_{perlin}$ the Perlin noise.

\subsection{Autoencoder Architecture}

We design a single-layer convolutional autoencoder for noise removal. The model consists of an encoder and a decoder, where the encoder compresses the input image into a latent representation. The encoder comprises a $3 \times 3$ convolutional layer with 32 filters followed by ReLU activation, and a $2 \times 2$ MaxPooling layer that reduces the spatial resolution by half.

The decoder reconstructs the compressed representation to the original image size. It begins with a $3 \times 3$ convolutional layer with 32 filters and ReLU activation, followed by a $2 \times 2$ UpSampling layer that restores the spatial resolution. The final layer uses 3 filters with sigmoid activation to normalize output values to the 0\textasciitilde1 range.

\subsection{Training Configuration}

The autoencoder is trained with input images of size $400 \times 400$, a learning rate of 0.0004, batch size of 8, and maximum epochs of 100. We use Adam as the optimizer and Mean Squared Error (MSE) as the loss function.

\begin{table}[h]
\centering
\caption{Training hyperparameters}
\begin{tabular}{ll}
\toprule
Parameter & Value \\
\midrule
Input size & $400 \times 400$ \\
Learning rate & 0.0004 \\
Batch size & 8 \\
Epochs & 100 \\
Optimizer & Adam \\
Loss function & MSE \\
\bottomrule
\end{tabular}
\label{tab:hyperparameters}
\end{table}

Training proceeds by minimizing the difference between original and reconstructed images, with the loss function defined as:

\begin{equation}
\mathcal{L}_{MSE} = \frac{1}{N}\sum_{i=1}^{N}(I_{original}^{(i)} - I_{reconstructed}^{(i)})^2
\end{equation}

where $N$ denotes the batch size, $I_{original}$ the original image, and $I_{reconstructed}$ the autoencoder-reconstructed image.

\subsection{Dataset Configuration}

We use vehicle-related images from the COCO dataset for our experiments. The training set consists of 18,135 images (80\%) from COCO train2017 vehicle-related images, while the validation set comprises the remaining 4,534 images (20\%). The test set consists of 999 vehicle-related images from COCO val2017 with adversarial noise applied.

The autoencoder is trained exclusively on clean images to learn the representation of unperturbed data. During validation and testing, adversarially perturbed images are used as inputs to evaluate the model's noise removal capability.

\subsection{Object Detection Evaluation}

We use YOLOv5 to evaluate object detection performance on reconstructed images. Performance metrics include COCO evaluation metrics: bbox mAP, bbox mAP@50, and bbox mAP@75. Evaluation is conducted under three conditions: original images without adversarial noise, adversarially perturbed images, and autoencoder-denoised images. This allows quantitative assessment of how effectively the proposed method recovers object detection performance degraded by adversarial attacks.

\section{Experiments}

\subsection{Experimental Setup}

Experiments are conducted on 999 vehicle-related images from the COCO val2017 dataset. We use a pretrained YOLOv5 as the object detection model and employ COCO standard metrics: bbox mAP, bbox mAP@50, and bbox mAP@75 for evaluation.

\subsection{Results}

Table \ref{tab:results} presents object detection performance under each condition.

\begin{table}[h]
\centering
\caption{Object detection performance under different conditions}
\begin{tabular}{lccc}
\toprule
Condition & bbox mAP & bbox mAP@50 & bbox mAP@75 \\
\midrule
Normal & 0.2890 & 0.4860 & 0.3000 \\
Adversarial (30\_30\_30\_2) & 0.1640 & 0.2780 & 0.1680 \\
Autoencoder & \textbf{0.1700} & \textbf{0.3080} & 0.1600 \\
\bottomrule
\end{tabular}
\label{tab:results}
\end{table}

\subsection{Analysis}

Experimental results demonstrate that Perlin noise-based adversarial attacks significantly degrade YOLOv5's object detection performance. Compared to original images, bbox mAP decreased by approximately 43.3\%, while bbox mAP@50 dropped by 42.8\% (from 0.4860 to 0.2780).

After applying the proposed autoencoder-based denoising, bbox mAP improved from 0.1640 to 0.1700 (3.7\% increase), and bbox mAP@50 increased from 0.2780 to 0.3080 (10.8\% improvement). These results indicate that the autoencoder effectively removes a portion of the adversarial noise, contributing to detection performance recovery.

However, bbox mAP@75 slightly decreased from 0.1680 to 0.1600. This is attributed to the loss of fine-grained details during the autoencoder's reconstruction process, which negatively impacts precise localization required at high IoU thresholds. Overall, the proposed method demonstrates the ability to partially recover object detection performance degraded by adversarial attacks.

\section{Discussion}

In this study, we evaluated whether autoencoder-based denoising can recover object detection performance degraded by adversarial attacks. Our experimental results demonstrate that the proposed method provides partial defense against Perlin noise-based adversarial attacks, suggesting that denoising can contribute to improving object detection robustness.

The Perlin noise-based adversarial attack reduced YOLOv5's bbox mAP from 0.2890 to 0.1640, a 43.3\% degradation. This performance drop is consistent with the effects of adversarial attacks reported in prior studies~\cite{xie2017adversarial, wei2018transferable}. Object detection tasks are particularly susceptible to severe performance degradation because both classification and localization are simultaneously affected, unlike image classification.

After applying the proposed autoencoder defense, bbox mAP improved from 0.1640 to 0.1700, and bbox mAP@50 increased from 0.2780 to 0.3080. These results demonstrate that the autoencoder effectively removes a portion of the adversarial noise, contributing to detection performance recovery. The larger improvement observed at the lower IoU threshold (mAP@50) suggests that the autoencoder is particularly effective at recovering coarse object localization. This can be interpreted similarly to the reformer concept in MagNet~\cite{meng2017magnet}, where the autoencoder projects adversarial examples toward the manifold of normal data.

Conversely, bbox mAP@75 slightly decreased from 0.1680 to 0.1600. This indicates that the precise localization required at high IoU thresholds is adversely affected by the loss of fine-grained details during the autoencoder's reconstruction process. When high-frequency information is lost during the compression-reconstruction process, accurate object boundary prediction becomes more difficult. As Liao \textit{et al}.~\cite{liao2018defense} noted, denoisers using only standard pixel-level losses face an inherent trade-off between adversarial noise removal and detail preservation.

\subsection{Limitations}

This study has several limitations. First, the representational capacity of a single-layer autoencoder is limited. Deeper architectures or designs incorporating skip connections, such as U-Net, may better balance detail preservation and noise removal. Nevertheless, the partial performance recovery achieved with our simple architecture validates the fundamental effectiveness of autoencoder-based defense.

Second, this study evaluated only a specific Perlin noise attack configuration. Additional evaluation on gradient-based attacks such as FGSM~\cite{goodfellow2014explaining}, PGD~\cite{madry2017towards}, and C\&W~\cite{carlini2017towards} might needed. Notably, defense effectiveness in white-box scenarios where attackers are aware of the autoencoder's presence was not addressed in this study.

Third, complete performance recovery relative to original images was not achieved. This may be attributed to the autoencoder's inability to remove all adversarial perturbations or the loss of detection-critical information during reconstruction. Combining our approach with randomization-based defense proposed by Xie \textit{et al}.~\cite{xie2017mitigating} or ensemble approaches with adversarial training may further enhance performance.

\subsection{Future Directions}

Future research could improve the balance between detail preservation and noise removal through deeper architectures such as U-Net or ResNet-based autoencoders. Additionally, applying high-level representation-based loss functions proposed in HGD~\cite{liao2018defense} to object detection tasks, directly leveraging the detection model's internal representations as guidance, warrants exploration. Extended evaluation across diverse attack methods and object detection models is necessary to validate the generalizability of the proposed defense mechanism.

\section{Conclusion}

In this study, we proposed and evaluated an autoencoder-based denoising defense to recover object detection performance degraded by adversarial noise. Experiments on vehicle images from the COCO dataset demonstrated that Perlin noise-based adversarial attacks reduced YOLOv5's bbox mAP by 43.3\%, while the proposed autoencoder defense achieved 3.7\% recovery in bbox mAP and 10.8\% improvement in bbox mAP@50. These results demonstrate that partial defense against adversarial attacks is achievable through preprocessing without model retraining. As an initial study quantitatively analyzing the effectiveness of autoencoder-based defense for object detection, this work can contribute to the development of practical adversarial defense systems through future research on more sophisticated denoiser architectures and diverse attack scenarios.

\bibliographystyle{ieeetr}
\bibliography{references}

\end{document}